\documentclass[reprint,prd,aps,twocolumn,superscriptaddress,eqsecnum,floatfix,preprintnumbers,amsmath,amssymb,
footinbib,longbibliography,citeautoscript]{revtex4-1}

\pdfoutput=1

\usepackage{graphicx}
\usepackage{epstopdf,epsfig}
\usepackage[scriptsize]{subfigure}
\usepackage[export]{adjustbox}
\usepackage{dcolumn}
\usepackage{bm}
\usepackage{marginnote}
\usepackage{changebar}
\usepackage[colorlinks]{hyperref}
\usepackage{wrapfig}
\usepackage{tcolorbox}

\usepackage{etoolbox}
\usepackage{ragged2e}
\patchcmd{\section}
  {\centering}
  {\justifying}
  {}
  {}

\usepackage{ccicons}

\newenvironment{WrapText}[1][r]
  {\wrapfigure{#1}{0.25\textwidth}\tcolorbox}
  {\endtcolorbox\endwrapfigure}

\setcitestyle{super}

\makeatletter
\def\@hangfrom@section#1#2#3{\@hangfrom{#1#2}#3}
\def\@hangfroms@section#1#2{#1#2}
\makeatother

\hypersetup{
    colorlinks=true,       
    linkcolor=blue,        
    citecolor=blue,        
    filecolor=black,       
    urlcolor=black,        
    bookmarks=false,
    pdffitwindow=true,
    pdfpagelayout=SinglePage,
    plainpages=false
}

\graphicspath{{figures/}}

\begin{document}

\title{Fluids in art: ``The water's language was a wondrous one, some narrative on a recurrent subject ...'' \\[0.25cm]
\textnormal{\normalsize Unexpected discoveries are made when classical paintings are analysed \\ on the subject of the faithful portrayal of ubiquitous fluid phenomena.}}

\author{R. Krechetnikov}
\affiliation{University of Alberta, Edmonton, Alberta, Canada T6G 2E1}

\date{\today}

\begin{abstract}
Artists spent a great deal of time studying anatomy for precise rendering of the human body as well as light, shadows, and perspective for convincing representation of the three-dimensional world. But in many paintings, they also had to depict fluids in their static and dynamic states -- a subject they could not study thoroughly, which led to a number of glaring misrepresentations or deliberate deceits.
\end{abstract}

\maketitle

\section{``Painters make their defeat certain by attempting to draw running water, which is a lustrous object in rapid motion.'' (John Ruskin \cite{Ruskin:1907})}

Since ancient times, artists studied \textit{proportions} -- the Greek sculptor Polykleitos (c. 450-420 B.C.), known for his ideal bronze Doryphoros and an influential Canon describing the proportions to be followed in sculpture -- as part of human \textit{anatomy} for realistic and convincing representations of their subject matter. In the treatise Della pittura (1435; ``On Painting'') Leon Battista Alberti urged painters to construct the human figure as it exists in nature, with musculature built on the skeleton, and only then draped in skin. For the same reason, it is believed that attempts to develop a system of \textit{perspective} began around the fifth century B.C. in Greece, to achieve illusionism in theatrical scenery. While first intuitive, these studies were also driven by a systematic look at the underpinning science -- the best examples including Leonardo da Vinci who performed dissections to uncover the structure of the human body resulting in the famous anatomical drawings, which are among the most significant achievements of the Renaissance. Altogether, \textit{on the artist side}, there has been an effort to convince the connoisseur of art. Can we say the same about rendering omnipresent fluid phenomena?

Similarly, artists had to represent still and flowing water since days of yore, e.g. the Ezekiel Mural ``The well of the wilderness: Moses gives water to the tribes'' at Dura Europos synagogue in III century or wall painting in The Tombs of Beni Hasan around 1900 B.C. with drawings of fishing and even humorously rescuing a human who fell in water \cite{Gombrich:1951}. Many centuries later, J. Ruskin -- an English writer, philosopher, art critic and polymath -- writes \cite{Ruskin:1984}: ``We want not to see water anatomized ... After the entire failure of all artists that ever lived before Turner in land and skies, we are prepared to find that they had not the least idea of water. When they thought they painted water, in fact, they were like `those happier children, sliding on dry ground,' and had not the chance of wetting a foot.'' In his opinion, J.M.W. Turner is ``the real Triton of the sea, as he was Titan of the earth''.

\begin{figure}[h!]
\centering
\includegraphics[width=3.25in]{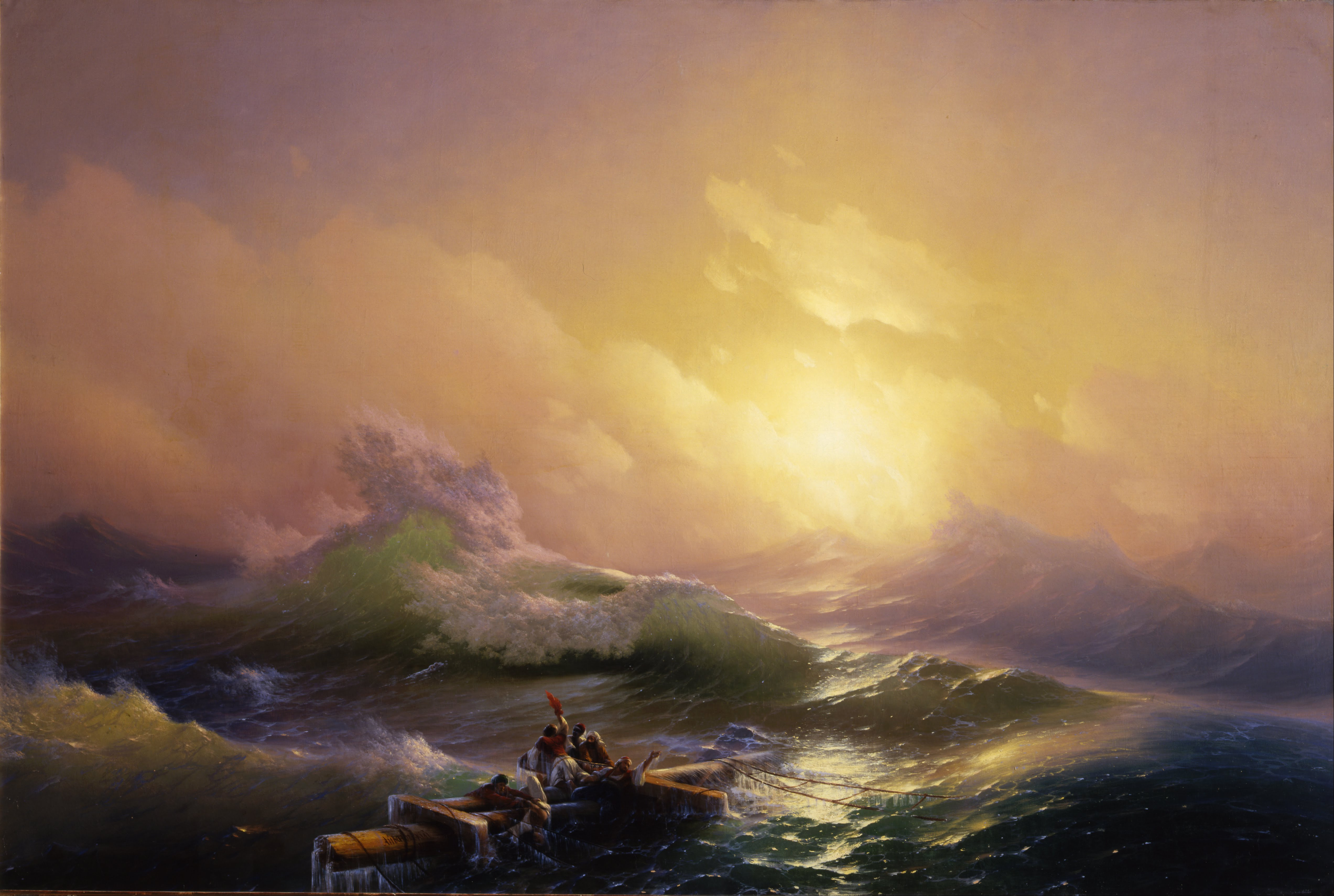}
\caption{``The Ninth Wave'' (1850) by Ivan Aivazovsky (1817-1900), The State Russian Museum, Mikhailovsky Palace. \ccPublicDomain}\label{fig:Aivazovsky}
\end{figure}
Seascapes, in fact, serve as a good example where artists are prone to mistakes and illustrate the fundamental reason for that: we will use one of them to introduce the reader to the subject that this article is dedicated to. In particular, breaking waves have always been the focus of maritime painters as they reflect the destructive power of nature with one of the most famous and analyzed \cite{Cartwright:2009,Dudley:2013,Ornes:2014} paintings of this sort being ``The Great Wave of Kanagawa'' by Hokusai. Another example illustrated in Fig.~\ref{fig:Aivazovsky} is Aivazovsky's depiction of ``The Ninth wave'' -- a colossal wave larger than all the others and the subject of legend among sailors -- formed in the middle of the sea about to crash on sailors survived after a shipwreck; however, the artist painted from the shore! Such a depicted wave can form only close to the coast, but not in the open sea, and is called a shore-break. As we now know from water wave theory, when a wave travels across the open sea, it gains speed, but when it reaches a shallow coastline, it begins to slow down due to the friction with the shallow bottom causing the wave to break. Later in his life, after painting about 3000 seascapes, in his autobiography Aivazovsky wrote: ``A painter, who just copies nature, becomes its slave, with his limbs tied up ... The living elements (of nature) are elusive to artist's brush.'' One of his latest paintings ``The Wave'' (1898) amazes with the details of the portrayed complexity of sea waves.

\section{``Imagine a child playing beside a pool of water, testing to see how close he can place his finger to the water surface'' (Cortat \& Miklavcic\cite{Cortat:2003})}

Let us start with depiction of still water, which presumably should be easier to render. For accuracy, we turn to academism, a favorite representative of which is William-Adolphe Bouguereau, who achieved a level of technical skill that was virtually unparalleled by his colleagues.
\begin{figure}
\centering
\includegraphics[width=3.25in]{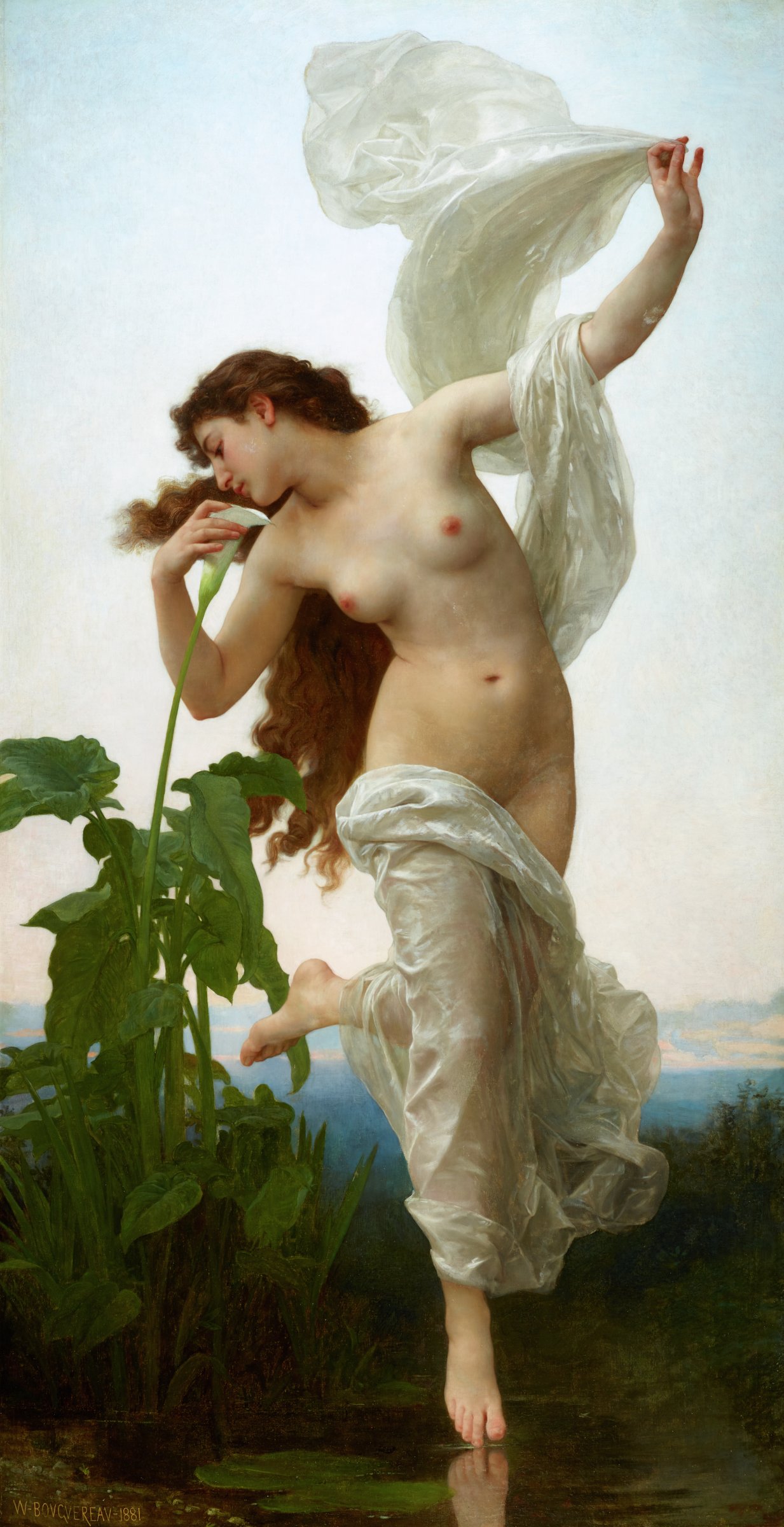}
\caption{``L'Aurore'' (Dawn) (1881) by William-Adolphe Bouguereau, Birmingham Museum of Art, Birmingham, Alabama, US. \ccPublicDomain}\label{fig:Bouguereau}
\end{figure}
In one of his most acclaimed and reverential works, ``L'Aurore'' in Fig.~\ref{fig:Bouguereau}, the water is like a mirror lightly perspired from a morning dew; as we can see, the reflection of a twin toe is rising to the surface to meet the other, and hence to touch the water surface. The artist's \textit{idea} behind this part of the painting was to convey the sensitive, unbearable lightness of being. Allegorically, for scientists this painting illustrates probing nature of water by carefully approaching it. Leaving aside that L'Aurore defies gravity, while water does not, the first look at the canvas brings up the question \cite{Cortat:2003} formulated in this section title and, under closer examination, shows impossibility of water to behave this way as there is no capillary rise where her toe seems to be touching the water surface. Obviously, the answer is dictated by the van der Waals interaction between finger/toe and water separated by the distance $h$ leading to the force per area $S$ characterized by the Hamaker constant $A_{H} \approx 10^{-20} \, \mathrm{J}$:
\begin{align}
\frac{F_{\mathrm{vdW}}}{S} = \frac{A_H}{6 \pi h^{3}}, \nonumber
\end{align}
which should be balanced by the gravity force $F_{g}$ acting on the bulge of water of height $h$ and of base area $S$ as well as the surface tension force $F_{\gamma} = \gamma/R$ tending to flatten the bulge, where the radius of curvature $R \approx S/(2 \pi h)$. As a result,
\begin{align}
\frac{F_{\mathrm{vdW}}}{S} = \frac{F_{g}}{S} + \frac{F_{\gamma}}{S} \ \Rightarrow \ \frac{A_H}{6 \pi h^{3}} = \rho g h + \frac{2 \pi \gamma}{S} h. \nonumber
\end{align}
Taking the area $S = 10^{-4} \, \mathrm{m^{2}}$ to be that of the finger/toe tip, cf. Fig.~\ref{fig:buldge} in Appendix, we can see that $F_{g}$ and $F_{\gamma}$ are of the same order and hence the minimal separation distance $h$ can be estimated via $h \approx \left(A_{H}/6 \pi \rho g\right)^{1/4} = 1 \, \mathrm{\mu m}$. However, this distance is much smaller than the normal tremor amplitude of human limbs \cite{Stiles:1976}, which starts at $O(100) \, \mathrm{\mu m}$ and increases to $O(1) \, \mathrm{cm}$ with time of continuous extension of a limb ($\sim 1 \, \mathrm{hr}$); presumably, the model had to pose much longer than that and hence touching water surface is unavoidable leading to capillary wetting. This vicious circle only means that Bouguereau's representation is physically impossible despite his realistic genre. In fact, his other painting ``The Wave'' suffers from a similar issue: the female figure appears to be completely dry despite the intense sea waves. Speaking of tremor in limbs, we can, albeit humorously, mention that the object of his painting is supposedly healthy, though science is capable of spotting neurological disorders in paintings as in the famous ``Christina's World'' by A. Wyeth \cite{Patterson:2017}. Same as how osteoarthritis, known to increase tremor, in the hands of Michelangelo Buonarroti was spotted from his portrait (c. 1535) by Jacopino del Conte \cite{Lazzeri:2016}.

\section{``Bacchus hath drowned more men than Neptune'' (Thomas Fuller)}

\begin{figure}[h!]
\centering
\includegraphics[width=3.25in]{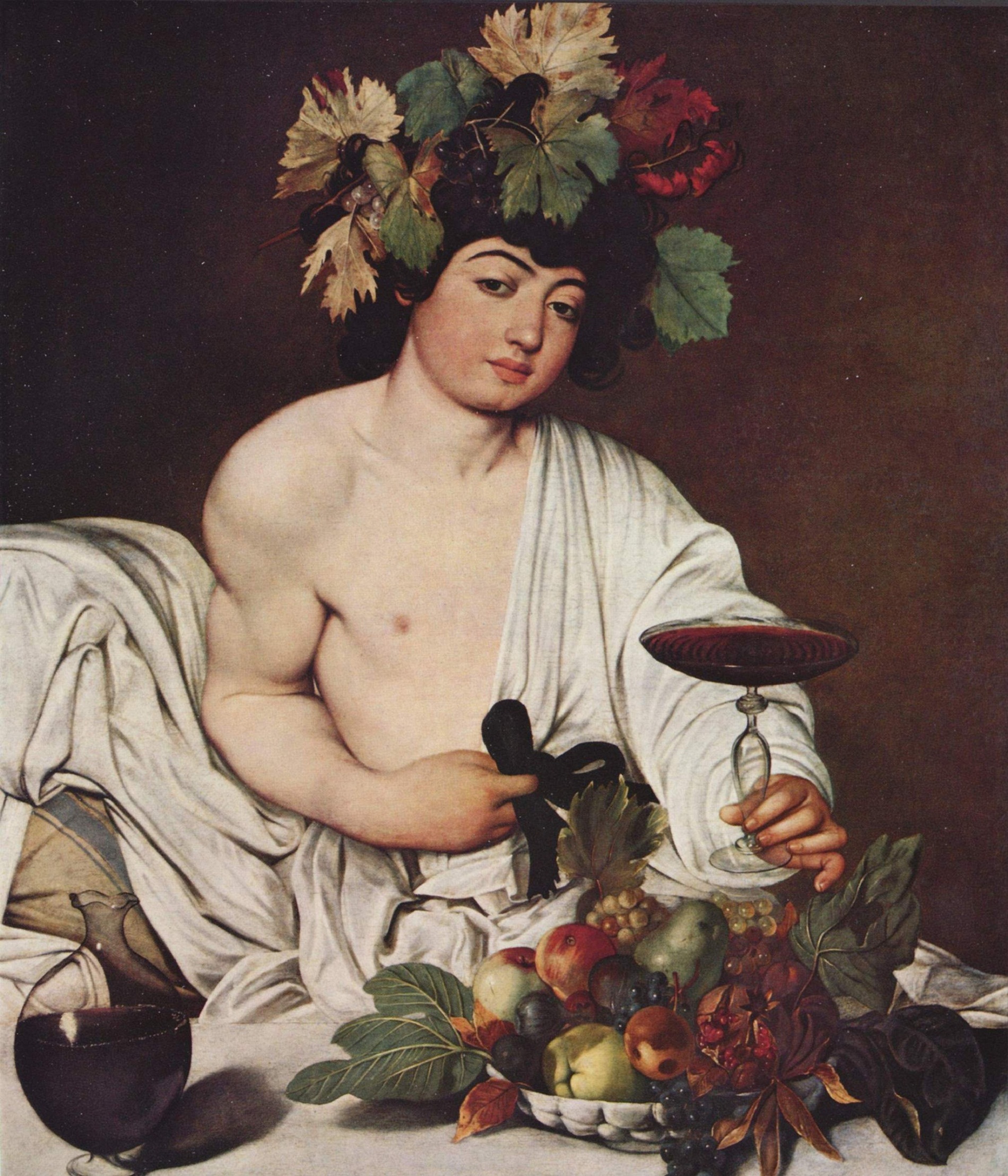}
\caption{``Bacchus'' (c. 1596) by Michelangelo Merisi da Caravaggio (1571-1610), Uffizi, Florence. \ccPublicDomain}\label{fig:Caravaggio}
\end{figure}
First look at the stark, dramatically lit super-realistic painting by Caravaggio, known for his uncunny level of detail, cf. Fig.~\ref{fig:Caravaggio}, suggests that the concentric waves in the wine glass -- likely concocted by the artist with the \textit{idea} to add some life to the painting as the flat wine surface is otherwise unappealing -- could be due to shaking hands of Bacchus, possibly caused by excessive drinking this deity is known for. Caravaggio himself was a heavy drinker, a womaniser, and of violent character which led him to committing three murders. This, nevertheless, did not prevent him from being one of the greatest artists of all times able to master an unprecedented accuracy. Thanks to the latter we can see that the wavelength $\lambda \approx 1 \, \mathrm{cm}$ is shorter than the depth of wine in the glass. Hence, we can determine the wave period from the standard relation for the gravity-driven waves, $\omega = (2 \pi g / \lambda)^{1/2}$, which gives circular frequency $\omega = \mathcal{O}(10^{2}) \, \mathrm{rad \, s^{-1}}$. Should the waves be generated from a point source -- e.g. by dipping and withdrawing a finger, which would be awkward to achieve as both hands of Bacchus are occupied -- then we know that the wavelength must be changing with the distance from the origin. Namely, according to the following expression for the wavelength $\lambda = 8 \pi r^{2} / (9 t^{2})$ for a given instant of time $t$ the wavelength must increase as $r^{2}$, which contradicts the painting, in which $\lambda$ does not change with the distance from the origin! Also, the waves are obviously shown for the time when they were able to propagate to the glass wall of radius $R$ at the wine level, that is from the wave propagation limit $R \sim g t^{2}/4$ we find $t > \left(4 R/g\right)^{1/2}$.

Hence, we conclude that the depicted waves, if true, must have a different origin rather than from a finger dipped in the glass. Another potential source of wave generation could be due to vibration of the wine glass itself, induced for example by hitting it against the table or some other object (or rubbing a moistened finger around the wineglass rim). Given the glass radius $R \approx 10 \, \mathrm{cm}$, thickness $a \approx 2 \, \mathrm{mm}$, density per unit area $\rho_{g} \approx 5 \, \mathrm{kg \, m^{-2}}$, Young modulus $E \approx 48 \, \mathrm{GPa}$, Poisson modulus $\nu = 0.22$, we can estimate the circular frequency of axisymmetric waves (formula is for a circular plate as the glass is almost flat):
\begin{align}
\omega_{g} = \frac{\lambda^{2}}{R^{2}} \left(\frac{D}{\rho_{g}}\right)^{1/2}, \ D = \frac{E a^{3}}{12 (1-\nu^{2})}, \nonumber
\end{align}
which yields for $\lambda^{2} = 10 - 10^{3}$: $\omega_{g} = 10^{3} - 10^{5} \, \mathrm{rad \, s^{-1}}$. Obviously, this frequency is incommensurate with the frequency of the observed water waves $\omega = \mathcal{O}(10^{2}) \, \mathrm{rad \, s^{-1}}$ calculated above, not to mention that these waves would have to propagate from the glass perimeter towards the center and hence wavelength must change and becomes shorter with a spike in the middle. Hence, this is not an option as well.

Finally, sloshing wine in the glass would not be able to induce symmetric waves of a uniform length, but would be a more convincing way to add life to the painting. Altogether, one can conclude that the depicted waves are inconsistent with the physics of water waves.

\section{``The Vikings' voyage was long and dangerous'' (Belikov \& Knyazeva \cite{Belikov:1973})}

\begin{figure}[h!]
\centering
\includegraphics[width=3.25in]{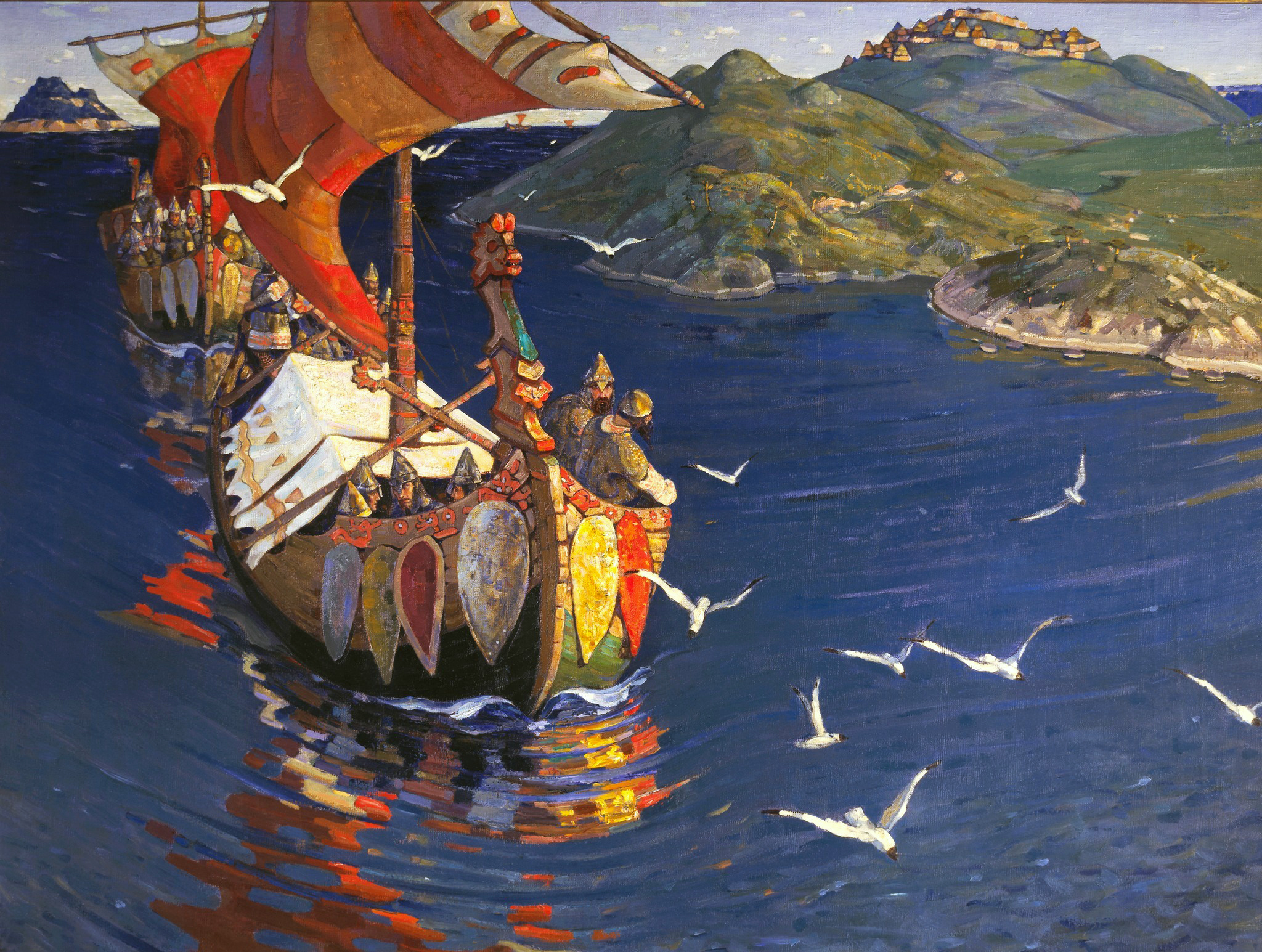}
\caption{``Merchants from Overseas'' (1901) is one of the most recognizable paintings by Nicholas Roerich (1874-1947), Tretyakov's gallery in Moscow. \ccPublicDomain}\label{fig:Roerich}
\end{figure}

\begin{WrapText}
\textbf{Sidebar}: His name is attached to Roerich Pact, which is an inter-American treaty protecting cultural objects from destruction for military purposes. Notably, Roerich also designed religious art for churches throughout both Russia and Ukraine.
\end{WrapText}
Laminar water waves on a larger scale have been depicted numerous times, but a notable artwork in this regard is by Nikolai Roerich on semi-fantastic motifs about the past Russia, shown in Fig.~\ref{fig:Roerich}. The canvas portrays a caravan of Viking ships sailed by merchants from distant countries. Obviously, they are not defenseless as on the rook sides there are large shields hanging, while inside the rooks people are shackled in iron armor.

The painting offers a number of phenomena suitable for basic estimates; the analysis here follows that by A. Stasenko \cite{Stasenko:1990}, though the conclusions are somewhat different. First, the waves in front of the boat are depicted as `supersonic' relative to the boat speed as they propagate ahead of the boat. By comparing with the head of Varangians, one can estimate their wavelength as $\lambda = 35 \, \mathrm{cm}$. Since, clearly, the wave propagation happens in the deep water regime, their speed can be calculated from the corresponding gravity-driven limit, $u_{g} = (g \lambda /2\pi)^{1/2} = 0.7 \, \mathrm{m \, s^{-1}}$. Second, the boat speed $v$ can be evaluated from the height $h$ of the bow wave based on the energy conservation for a mass of water $m$ moving with velocity $v_{\perp} = v \sin{\alpha}$ towards to the boat wall at the bow of angle $2 \alpha$: $m g h \le m v_{\perp}^{2}/2$. Estimating the bow wave height from the painting to be $h \approx 25 \, \mathrm{cm}$ and the bow angle $\alpha \approx \frac{\pi}{2}$, we conclude that $v \ge \left(2 g h\right)^{1/2}/\sin{\alpha} \approx 2.3 \, \mathrm{m \, s^{-1}}$. Obviously, there is a contradiction to the observation made earlier that the water waves are `supersonic'. Given that the reflections on the waves ahead of the boat seem to be quite realistic suggesting that they were copied from a real model, one might guess that it is the wave height at the bow, which was misrepresented for an artistic purpose. The conundrum resolution depends on which part of the painting you trust. The artist's \textit{idea} with waves propagating ahead of the boat was to project forceful arrival of Vikings. However, Roerich cannot be excused for making the phenomena unrealistic since linear water wave theory has already existed by that time due to George Biddell Airy (1841).

\section{``What care I that the virtue of some sixteen-year-old maid was the price for Ingres' La Source? That the model died of drink and disease in the hospital is nothing when compared with the essential that I should have La Source, that exquisite dream of innocence.'' (George Moore)}

\begin{figure}[h!]
\centering
\includegraphics[width=3.25in]{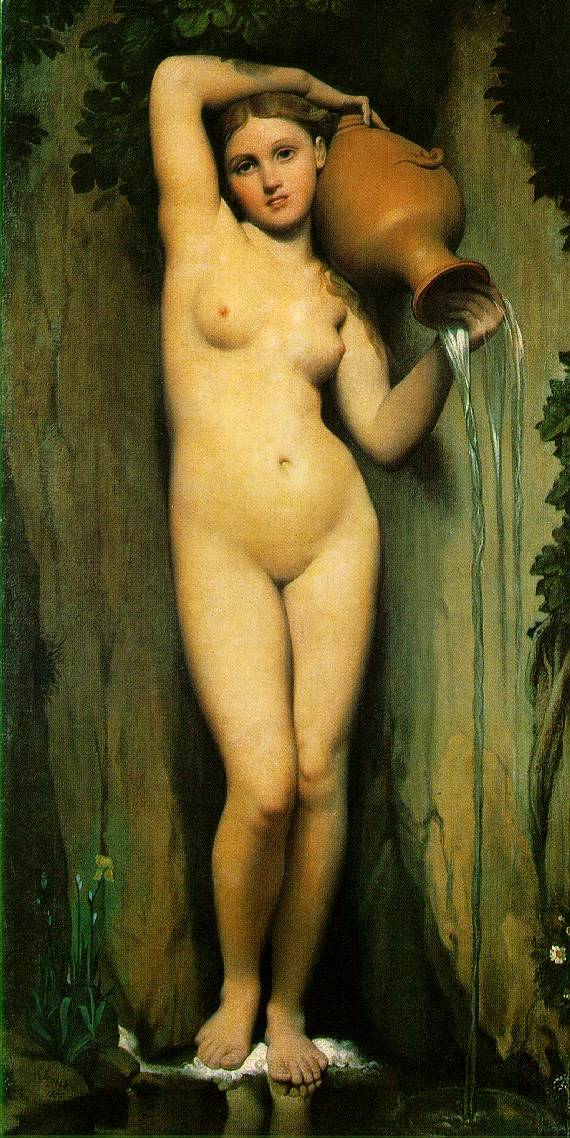}
\caption{``La Source'' (1820-1856) by Jean-Auguste-Dominique Ingres (1780-1867), Mus\'{e}e d'Orsay, Paris. \ccPublicDomain}\label{fig:Ingres}
\end{figure}
Waves are ubiquitous manifestations of fluid dynamics not only on horizontal water surfaces discussed above, but also on flowing jets and streams, as we can see in Ingres' painting, cf. Fig.~\ref{fig:Ingres}. A reader might say: ``Enough nudes!'' Gustav Klimt would have objected ``All art is erotic''. It just happen that nudes constitute a substantial part of art, especially from that period. And there is a good reason for that, i.e. to show people outside of their historical context. The work on ``La Source'' was begun in Florence around 1820 and completed in Paris around 1856, when he was seventy-six years old, already being the famous president of the \'{E}cole des Beaux-Arts. A water spring, which in French is \textit{la source}, is represented here in the guise of a girl pouring water from a jug; in classical literature, a source is also sacred to the Muses of poetic inspiration. The young daughter of Ingres' concierge served as a model for the painting.

Despite the long period of work on this painting, there are a few issues which become apparent due to meticulous attention to details provided by the artist. First, by noting that there is a well-defined circle -- an expanding wave on the water pond surface -- from general theory of gravity-driven waves we know that its radius is $r = g (t_{d}-t_{i})^{2}/4$ where $t_{d}$ is the movement of depiction and $t_{i}$ the moment of the first jet impact on the pond, which gives us $t_{d} - t_{i} \approx 0.2 \, \mathrm{s}$, i.e. $t_{d} \approx 0.6 \, \mathrm{s}$. The moment $t_{i}$ follows from the basic motion under gravity giving $t_{i} \approx 0.4 \, \mathrm{s}$; the corresponding jet velocity is $V_{i} \approx 5 \, \mathrm{m/s}$ after it travels $\approx 1 \, \mathrm{m}$. Finally, the water exit velocity $V_{0}$ from the pitcher at the moment of depiction follows from Bernoulli's equation $V_{0} = (2 g h)^{1/2}$, where for $h$ we estimate $h = 0.2 \, \mathrm{m}$ thus giving $V_{0} \approx 2 \, \mathrm{m/s}$.

A close look reveals that there are three jets: one on the left of the nude's hand (viewer's point of view) reaches the pond, while the other two enveloping her hand traveled only about $0.5 \, \mathrm{m}$. The only explanation for the different lengths of these jets is that the pitcher was tilted more some time after initial pouring started in order to overflow the obstruction due to the hand. The jet dynamics raises even more troubling issues. For that notice that the left jet diameter $d_{1} \approx 1.5 \, \mathrm{cm}$ based on the comparison with the model's iris. At the exit from the pitcher the same jet is approximately elliptic with two major axes $a$ and $b$, so that the equivalent radius $d_{e} = (a b)^{1/2} \approx 2 \, \mathrm{cm}$ at the top of the jet. The axis-switching wavelength varies approximately from $\lambda_{1} \approx 5 \, \mathrm{cm}$ at the top of the jet to $\lambda_{1} \approx 8-10 \, \mathrm{cm}$ at the bottom. The axis-switching of elliptic jets studied by Lord Rayleigh \cite{Rayleigh:1879} and recently improved to account for gravity \cite{Ma:2020} yields for the circular frequency of axis-switching:
\begin{align}
\omega = \omega_{0} \left(1 + \frac{2 g z}{V_{0}^{2}}\right)^{1/4}, \ \omega_{0} = 2 \sqrt{3} \left(\frac{\sigma}{\rho}\right)^{1/2} d_{e}^{-3/2}, \nonumber
\end{align}
where $d_{e}$ is an equivalent diameter of the jet, $z$ is the distance from the exit from pitcher, and $V_{0}$ the jet exit velocity. The wavelength is then found from $\lambda(z) = 2 \pi V(z) / \omega(z)$. Plugging in numbers shows that the instability wavelength varies from $\lambda \approx 0.15 \, \mathrm{m}$ at the top to $\approx 0.5 \, \mathrm{m}$ at the bottom, which agrees with the laboratory measurements \cite{Ma:2020}. Thus, while Ingres correctly captured the qualitative feature that wavelength of axis-switching must increase along the jet travel direction, the depicted wavelength is several times shorter! This indicates that he likely used a scale-down experiment as a model for his painting. More importantly, the Weber $We = \frac{\rho V^{2} d}{\sigma}$ and Reynolds $Re = \frac{\rho V d}{\mu}$ numbers for the jet on the left for $d$ varying from $d_{1}$ to $d_{e}$:
\begin{align}
We \approx (1-4) \cdot 10^{5}, \ Re \approx 5 \cdot 10^{6}, \nonumber
\end{align}
imply that the jet is far in the unstable regime accompanied by jet breakup and fish-bone structures! Presumably, Ingres knew that the jet breaks up into droplets, but likely found it unappealing. Notably, its namesake by Gustave Courbet (1862) does not suffer from incorrect depiction of water jet.

\section{``We are meaner than flies; flies have their virtues, we are nothing but bubbles.'' (\textit{Satyricon}, Petronius, 1st century C.E.)}

The symbolism of \textit{Homo bulla} (man is a bubble) was already proverbial by the first century B.C. (``De Re Rustica'', Varro, 116-27 B.C.). In the baroque and rococo eras soap bubbles conveyed allegorically the fleeting nature of earthly pleasures and to remind the viewers of the fragility of the human existence \cite{Behroozi:2008,Hosseini:2020}. As they have been a source of inspiration to artists, so for scientists as well: ``Make a soap bubble and observe it; you could spend a whole life studying it,'' (Lord Kelvin).

\begin{figure}
\centering
\includegraphics[width=3.25in]{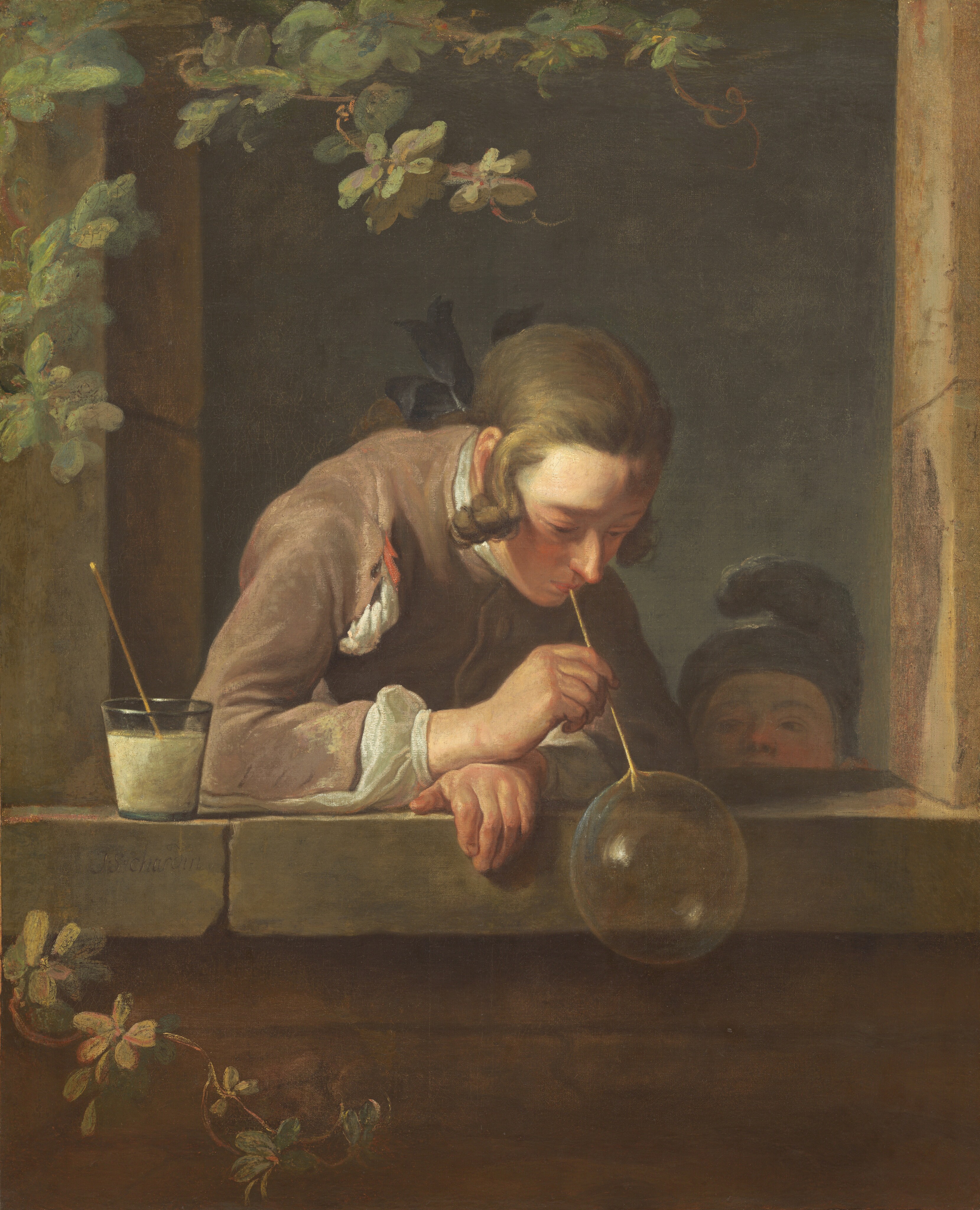}
\caption{``Soap bubbles'' (Les Bulles de Savon) (1733-1734) by Jean-Baptiste-Siméon Chardin, Metropolitan Museum of Art, New York City. A version of it was exhibited at the Paris Salon of 1739, where it was widely acclaimed. \ccPublicDomain}\label{fig:Chardin}
\end{figure}
Jean-Sim\'{e}on Chardin\footnote{Besides him, soap bubbles appear in works of many other artists in various countries including contemporaries Gerrit Dou, Karel Dujardin, Charles Van Loo, Jan Steen, Edward Wilson, John Everett Millais, Thomas Couture, Edouard Manet, Hans Heyerdahl, to name a few. Not so long time ago, the art world was shaken with the stolen painting ``Child with a Soap Bubble'' in 1999, which is believed to belong to Rembrandt. The painting was returned to the municipal museum of Draguignan in March, 2014.} transformed the genre of still-life painting and contributed to making France the origin of a new dimension of critical understanding in art, a new feeling of life, in which individual intelligence was opposed to wealth and power. Compared to still life painters from the previous century such as Frans Hals, he added something more profound such a smile shedding light on hypocrisy, the mood of quiet absorption, so familiar to scientists, that can lead to new knowledge. Chardin's picture in Fig.~\ref{fig:Chardin} definitely hints at the transient nature of life. Such images resonate in the context of the period's growing interest in the experience of childhood, culminating in Jean-Jacques Rousseau's influential treatise ``\'{E}mile'', or ``on Education'' (1762) about the right way to raise children. Rousseau himself, however, betrayed his own and sent them to orphanage immediately upon birth without ever knowing or seeing them. The price to pay for genius?

First, from the painting in figure~\ref{fig:Chardin} one learns that split straw was used to create soap bubbles. One may ask why is the bubble conspicuously elongated and why is it not hanging much down due to gravity compared to paintings of other artists? Careful examination of the canvas suggests that with the average radius $R \approx 10 \, \mathrm{cm}$ the main axes of the ellipsoid bubble differ by about $1 \, \mathrm{cm}$, i.e. $\delta R \equiv R_{\mathrm{max}}-R_{\mathrm{min}} \approx 1 \, \mathrm{cm}$. Also, the bubble main axis is at about $\theta_{0}=\frac{\pi}{6}$ from the vertical axis and deviates from that of the straw by about $\delta \theta \approx 0.1 \, \mathrm{rad}$. Hence, the bubble volume $V_{b} \approx 4 \cdot 10^{-3} \, \mathrm{m^{3}}$. We also need the soap film thickness. Since we cannot see colors on the bubble, it must be above the visible light wavelength. At the same time, given inefficiency of the old method to produce soap bubbles, it is likely that the depicted soap bubble is close to bursting, so we can take for its thickness $h = 1 \, \mathrm{\mu m}$, which gives for the volume of water in the soap film $V_{\mathrm{SF}} = 4 \pi R^{2} h \approx 10^{-7} \, \mathrm{m^{3}}$ and mass $m_{\mathrm{SF}} \approx 0.1 \, \mathrm{g}$. Also, the soap was not as efficient as the one produced commercially nowadays such as sodium dodecyl sulfate, which lowers surface tension of water from $0.07 \, \mathrm{N \, m^{-1}}$ to about $0.03 \, \mathrm{N \, m^{-1}}$; hence we take for the surface tension of the soap solution in the painting $\sigma = 0.05 \, \mathrm{N \, m^{-1}}$.

Next, applying Bernouilli's equation to the streamline impinging the front of the soap bubble, cf. Fig.~\ref{fig:bubble} in Appendix, we can relate the stagnation pressure $p_{s}$ at the umbilic of the bubble with that at the exit of the straw $p_{e}$, i.e. $p_{s} = p_{e} + \frac{1}{2} \rho u_{e}^{2}$. Also, from the Young-Laplace law we know the pressure jump across the bubble interface at the stagnation point $s$:
\begin{align}
p_{s} - p_{0} = \frac{4 \sigma}{R - \frac{3}{2}\delta R} \approx \frac{4 \sigma}{R}\left(1 + \frac{3}{2}\frac{\delta R}{R}\right). \nonumber
\end{align}
Taking into account that should there be no flow, $u=0$, the bubble would be round, $p_{e} - p_{0} = \frac{4\sigma}{R}$. Altogether, this gives the expression for the ejection velocity $u^{2} = \frac{12 \sigma}{\rho} \frac{\delta R}{R^{2}}$, which yields $u \approx 1 \, \mathrm{m \, s^{-1}}$.

Finally, considering the horizontal ($x$) and vertical ($y$) components of the force acting on the bubble,
\begin{align}
F_{x} \approx p_{s} S \sin{\theta_{0}}, \ F_{y} \approx p_{s} S \cos{\theta_{0}} + \delta m \, g, \nonumber
\end{align}
where $S = \pi R^{2}$ is the cross-section of the bubble and $\delta m \, g = \delta \rho \, V_{b} \, g$ is the buoyancy force due to the density difference between air inside and outside the bubble, we get
\begin{align}
\tan{(\theta_{0} + \delta\theta)} = \frac{F_{x}}{F_{y}} \ \Rightarrow \ \delta \rho = \frac{2 \, \delta \theta \, p_{s} S}{g V_{b}} \approx 10^{-2} \, \mathrm{kg \, m^{-3}},  \nonumber
\end{align}
i.e. the air is a bit heavier inside the bubble, which is due to exhaled microdroplets of water (evaporated water, not in the form of microdroplets, in fact makes air lighter!) and $\mathrm{CO_{2}}$. Of course, the bubble equilibrium could be easily swayed by the convecting air, which was not taken into account. The consistency of the calculations with the artist depiction suggests that there was no significant flow of air at the moment of observations, which is perhaps true for the sunny afternoon in French Riviera. Chardin produced quite a number of paintings with soap bubbles, which means that he inevitably studied them well and hence depicted accurately.

\section{Summary}

One might suppose that the difference between the artist and the scientist is that the artist seeks beauty, while the scientist seeks truth. While S. Chandrasekhar argued \cite{Chandrasekhar:1987} that this is not the case, in this work we sought both truth and beauty in artwork. The title of this article is (a translation by A. Shafarenko) from poetry of Arseny Tarkovsky -- a Soviet and Russian poet, who was arrested and sentenced to execution for publishing a poem which contained an acrostic about Lenin, but managed to escape -- and encapsulates the richness of the language one may speak with fluid phenomena.

While fluid phenomena themselves are a great source of inspiration for artists \cite{Dyke:1982,Zabusky:2015}, as we saw from the analyzed paintings, artists often used fluids as a conceptual means to communicate certain emotions: Aivazovsky's ``The Ninth Wave'' (Late Romanticism) -- destructive power of nature, Roerich's ``Merchants from Overseas'' (Russian realism) -- forcefulness of incoming Vikings, Ingres' ``La Source'' (Neoclassicism) -- poetic inspiration, Chardin's ``Soap bubbles'' (Rococo) -- fragility of human existence, Bouguerau's ``L'Aurore'' (Academism) -- unbearable lightness of being, Caravaggio's ``Bacchus'' (Baroque) -- the liveliness of wine. While in some cases, such as ``Bacchus'', one can say that the artist corrected nature, in most of fluid phenomena misrepresentations the mistakes were genuine based on the lack of understanding of fluid phenomena; otherwise it would go against aspirations to portray the world truthfully as discussed at the beginning of the article. However, all the contortions could not mar the harmonious beauty of the canvas. Besides debunking the scientifically troubled renderings of fluid phenomena, we also tried to highlight sometimes tragic background or hypocritical motivation: the deceptive nature of aesthetically valued artwork goes well beyond picturing fluid dynamics.

\section{Instead of an afterword}

\begin{figure}[h!]
\centering
\includegraphics[height=1.55in]{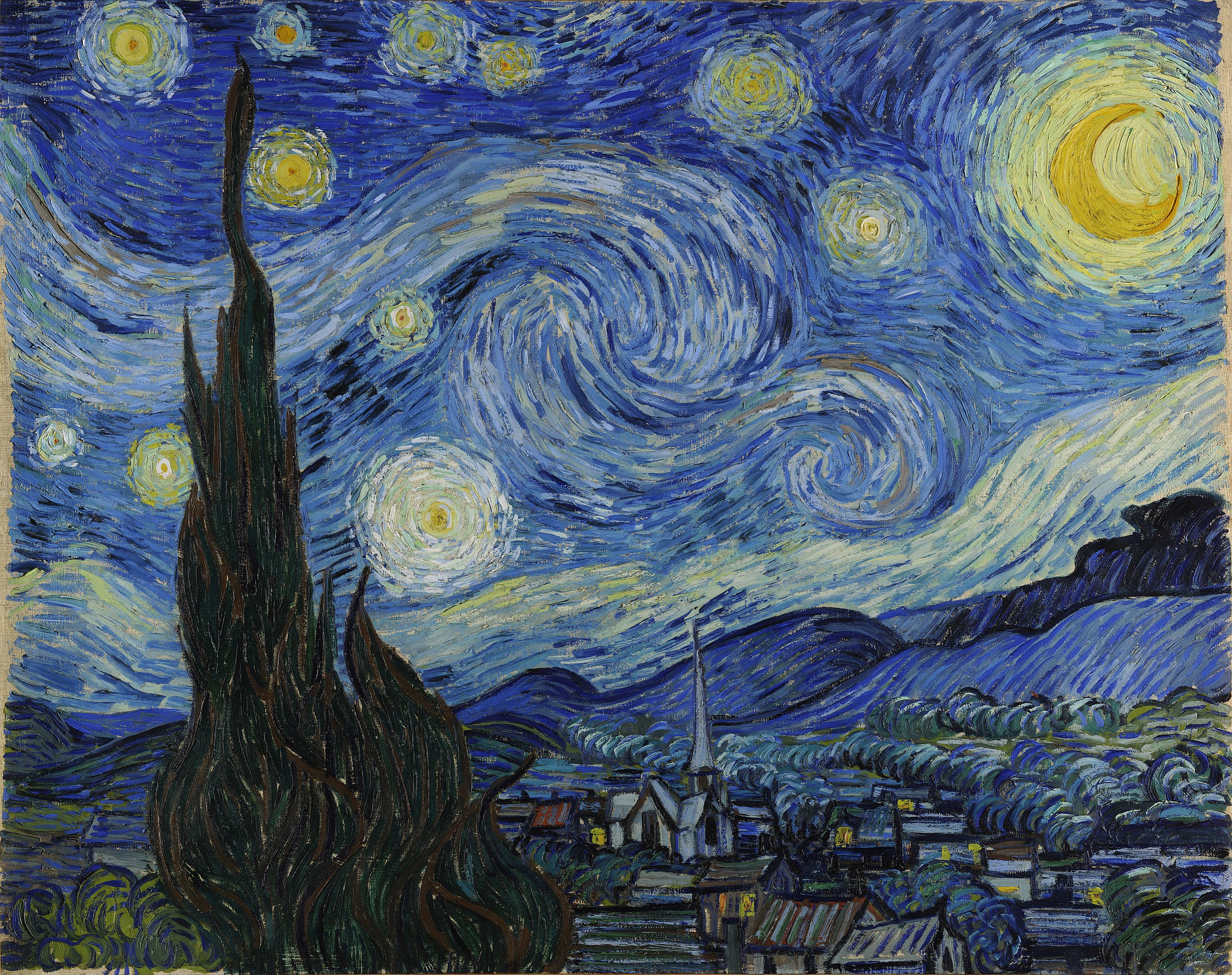}
\includegraphics[height=1.55in]{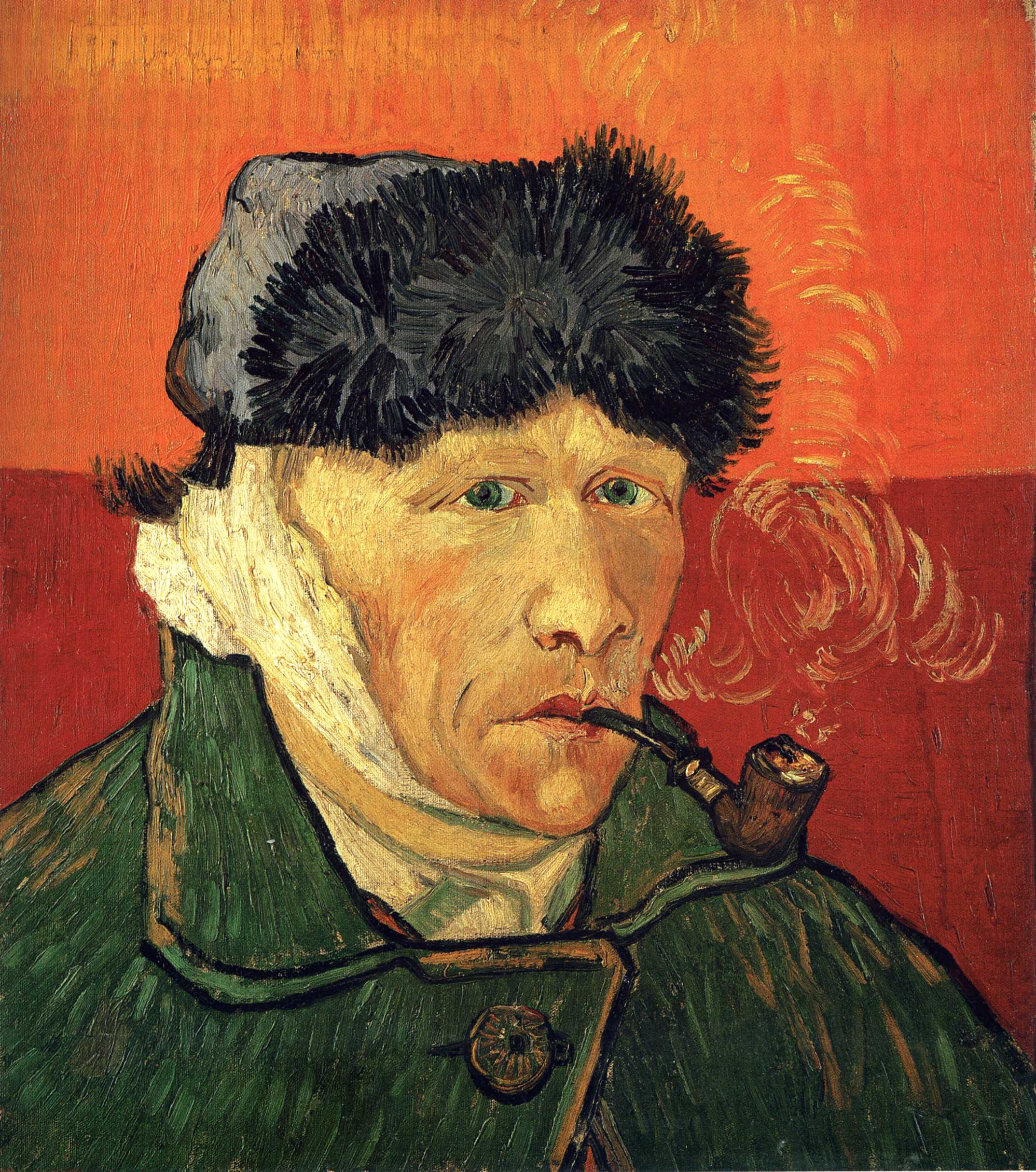}
\caption{(left) ``The Starry Night'' (1889), Museum of Modern Art, New York City. (right) ``Self-portrait with Bandaged Ear and Pipe'' (1889), Courtauld Institute, London by Vincent van Gogh (1853-1890). \ccPublicDomain} \label{fig:Gogh}
\end{figure}

To reinforce this point of view, and to append our telegrammatic excursion in the wide range of epochs and movements, we invite the reader to ponder over a few more paintings from impressionism and post-impressionism, to contrast with, say, academism. In fact, the Impressionists had the gall to open their first exhibition in 1874 in opposition to ingrained academic favorites like Bouguereau, who was then at the height of his career as a Salon master. Turning to Van Gogh, whose expressive exaggerations are perhaps most clearly seen in his last works, painted in 1889, during his stay in a psychiatric facility hospital in Saint-R\'{e}my, near Arles. Both paintings are shown in Fig.~\ref{fig:Gogh} and fluid depictions there aim to deliver vivacity of the sky in van Gogh's ``The Starry Night'' as well as human addiction in his ``Self-portrait with bandaged ear and pipe''.``The Starry Night'', which was a view from Vincent's window, distorted to express the artist's existential turmoils. In both paintings van Gogh tried to capture the essence of the apparently turbulent fluid flows, which we formally call coherent structures nowadays. Impressionism and coherent structures, discovered in science only relatively recently \cite{Brown:1974}, go hand-in-hand despite that the idea of depicting complex phenomena via the key features goes back at least to Leonardo da Vinci, who was engaged with fluid flows throughout his life \cite{MacCurdy:1938,Macagno:2000}. It has been noticed by fluid mechanicians \cite{Taylor:1974,Lumley:1992} that Leonardo ``seems to be thinking about ways of separating flow into steady and turbulent components'' and presage the concept of coherent structures \cite{Gad-el-hak:1998,Marusic:2021} as the words eddies and eddying motions in turbulenza percolate throughout Leonardo's treatise of liquid flows\footnote{``Observe the motion of the surface of the water, which resembles that of hair, which has two motions, of which one is caused by the weight of the hair, the other by the direction of the curls; thus the water has eddying motions, one part of which is due to the principal current, the other to the random and reverse motion.'' (Leonardo, c. 1500)}.

One would expect that impressionists, including van Gogh, could have learned not only from da Vinci, but also benefited from the contemporary science \cite{Guebhard:1881}. The question to ask is if in ``The Starry Night'' the artist respected the regular structure of van Karman street by arbitrarily terminating the von Karman vortices or, may be, he captured the transient stage of the street formation. Compare this to the famous photographs of the Kelvin-Helmholtz instability, e.g. in ``The Album of Fluid Motion''\cite{Dyke:1982}. Another canvas, cf. Fig.~\ref{fig:Gogh}(b), is his self-portrait.

Leaving aside that in reality he cut his left ear, the smoke coming out of the pipe is able to develop fanciful large scale instabilities (vortices) a short distance from the pipe: compare to the smoke plume photographed by Werner Wolff -- Black Star \cite{Corrsin:1961}. In the latter case, the combustion and release of hot, opaque compounds form a column of smoke, warmer than the surrounding air, which develops upward, forming a laminar flow. At some point, however, free convection weakens as the smoke column cools down and thus cannot maintain its stability by advecting irregularly in the surrounding air forming eddies.

\begin{figure}[h!]
\centering
\includegraphics[width=3.25in]{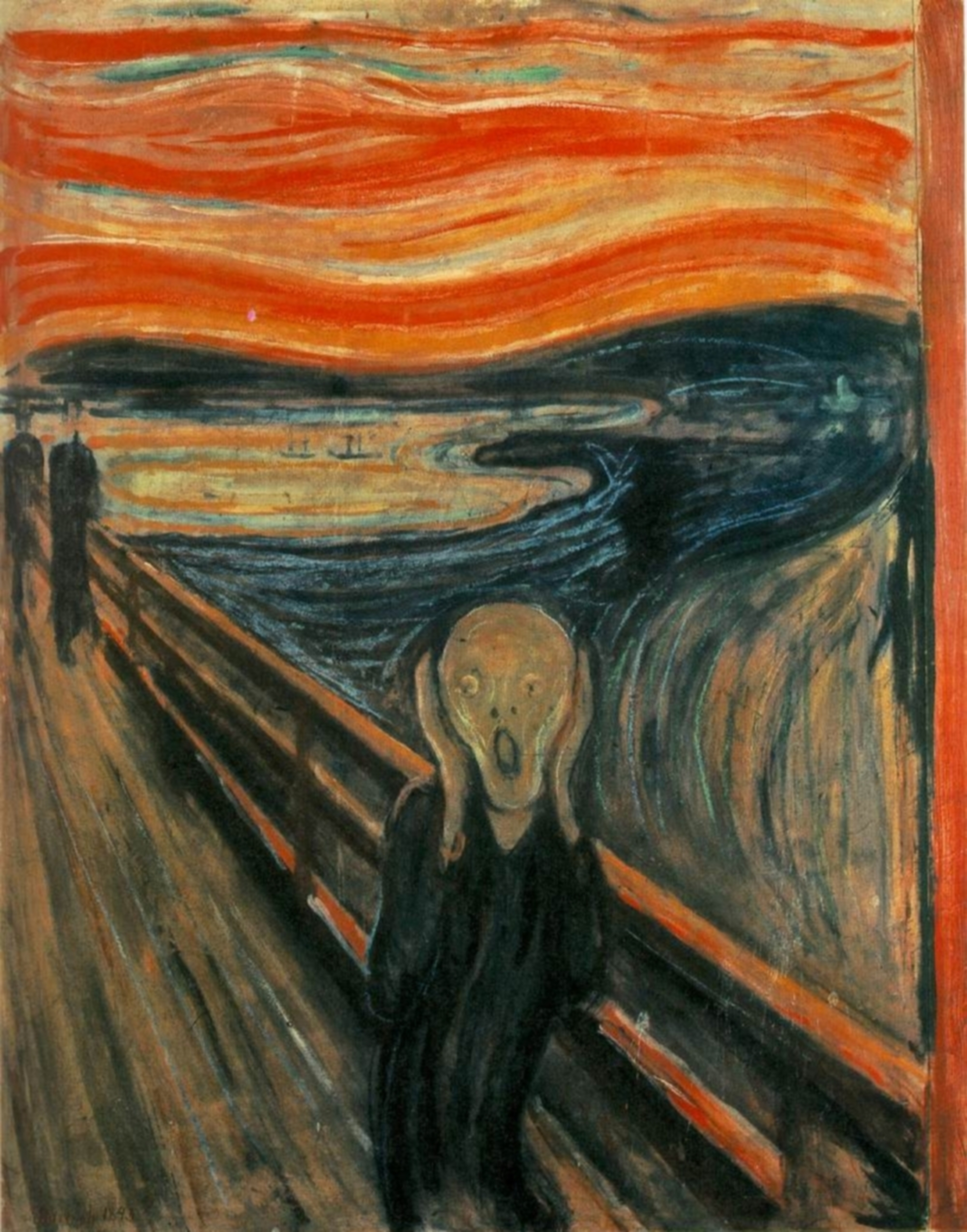}
\caption{``The Scream'' (1893) by Edvard Munch (1863-1944), National Gallery of Norway. \ccPublicDomain}\label{fig:Munch}
\end{figure}
Echoed van Gogh's mood is the final powerful canvas ``The Scream'' of our discussion shown in Fig.~\ref{fig:Munch}, symbolizing the anxiety of the human condition or distress based on a personal experience of the artist described in his autobiographical prose poem\footnote{
I was walking along the road with two friends.\\
The sun was setting. I felt a breath of melancholy\\
--Suddenly the sky turned blood-red. I stopped,\\
and leaned against the railing, deathly tired--\\
looking out across the flaming clouds that hung\\
like blood and a sword over the blue-black fjord\\
and town. My friends walked on--I stood there,\\
trembling with fear. And I sensed a great, infinite\\
scream pass through nature.} However, the figure in the painting is not just ``sensing'' but appears to produce the scream with the reverberation of the sound throughout the vast landscape suggested by the long curving strokes.

While we see tall ships in the background, it is an invisible fluid dynamics -- acoustics -- which is our focus now. It is believed to be inspired by the 1883 Krakatoa eruption, which exploded with such a force that it destroyed its island and atoll, releasing 20 million tons of ash into the atmosphere that it made sunsets reddish for years. It also produced the loudest ``scream'' ever recorded: measuring instruments 160 km (100 mi) away from Krakatoa picked up an immense 172 decibels, quite loud considering the distance. For comparison, humans start to experience acute pain at 125 decibels.

Given the inverse square law for the sound pressure $L$ reduction with distance $R$ from a source
\begin{align}
L_{2} - L_{1} = - 20 \log{\frac{R_{2}}{R_{1}}}, \nonumber
\end{align}
we can find the range of sound level $L_{1}$ of the screamer given that the two strangers behind the screamer do not cover their ears. Taking $R_{1} = 0.1 \, \mathrm{m}$ as the distance from his mouth to ear and $R_{2} = 10 \, \mathrm{m}$, we find $L_{2}-L_{1} = -40 \, \mathrm{db}$, i.e. since $\max{L_{2}} = 125 \, \mathrm{db}$, we get $L_{1} \le 165 \, \mathrm{db}$. If one naturally assumes that the screamer covers his ears due to acute pain, then $\min{L_{1}} = 125 \, \mathrm{db}$ and hence at the strangers location $85 \, \mathrm{db} \le L_{2} \le 125 \, \mathrm{db}$. This is a believable range as the world's loudest shouting record with $121.7 \, \mathrm{db}$ (equivalent to a jet engine) was produced by Irish teacher Annalisa Flanagan.

\begin{acknowledgments}
The author is grateful to Professor A. Stasenko at my Alma Mater (Moscow Institute of Physics and Technology), whose article \cite{Stasenko:1990}, originally published in Russian in 1973, sparked a life-long interest in the subject of depicting fluid phenomena in art -- the illustrations of which have been used by the author in fluid mechanics classes both at the University of California at Santa Barbara and University of Alberta. Special thanks to Dr. A. Zelnikov for many stimulating discussions as well as Dr. A. Herczynski for his feedback on an earlier version of this manuscript.
\end{acknowledgments}

\appendix*

\section{Auxiliary figures}

\begin{figure}[h!]
\centering
\includegraphics[width=1.75in]{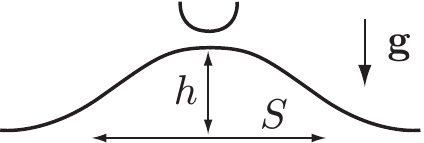}
\caption{On critical distance.}\label{fig:buldge}
\end{figure}
\begin{figure}[h!]
\centering
\includegraphics[width=1.5in]{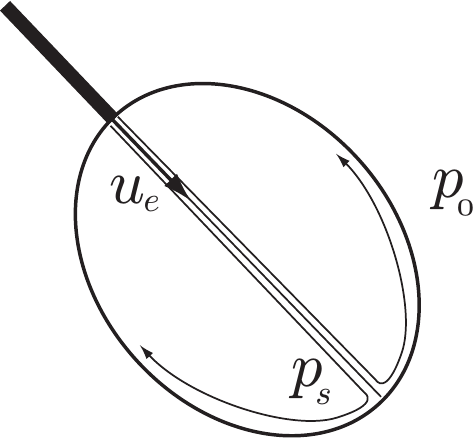}
\caption{Air flow in the bubble.}\label{fig:bubble}
\end{figure}

\bibliography{references}

\end{document}